\documentclass[12pt,showpacs]{revtex4}
\usepackage{graphicx}
\usepackage{amsmath}
\usepackage{amssymb}
\usepackage{color}

\begin{document}
\title{Revealing the role of electron correlation in sequential double ionization}
\author{Pengfei Lan$^{1,3}$, Yuming Zhou$^1$, Adrian N. Pfeiffer$^2 \footnote{Present address:
Institut f\"{u}r Optik und Quantenelektronik,
Friedrich-Schiller-Universit\"{a}t, Max-Wien-Platz 1, 07743 Jena,
Germany.}$, Qingbin Zhang$^1$, Peixiang
Lu$^1\footnote{Corresponding author:
lupeixiang@mail.hust.edu.cn}$}
\author{Katsumi Midorikawa$^3$}
\affiliation{$^1$School of Physics and Wuhan National Laboratory for Optoelectronics, Huazhong University of Science and Technology,
Wuhan 430074, P. R. China
\\ $^2$ Physics Department, ETH Zurich, 8093 Zurich, Switzerland
\\ $^3$ Extreme Photonics Research Group, RIKEN Center for Advanced Photonics, RIKEN, 2-1 Hirosawa, Wako, Saitama 351-0198, Japan}
\date{\today}

\begin{abstract}
The experimental observations of sequential double ionization
(SDI) of Ar [A. N. Pfeiffer {\it et al.}, Nature Phys. {\bf 7},
428 (2011)], such as the four-peak momentum distribution and the
ionization time of the first and second electrons, are
investigated and reproduced with a quantum model by including and
excluding the $e$-$e$ correlation effect. Based on the comparison
of experiment and simulation, the role of $e$-$e$ correlation in
SDI is discussed. It is shown that the inclusion of $e$-$e$
correlation is necessary to reproduce the momentum
distribution of electrons.
\end{abstract} 

\maketitle

Tunneling ionization of atoms, molecules and semiconductors
exposed to strong laser fields is one of the most fundamental
quantum processes and has been of significant interest over the
past several decades \cite{review}. Theories \cite{PPT,ADK} have
been well established for single electron ionization based on the
Keldysh frame \cite{Keldysh}. By numerically solving the
time-dependent Schr\"{o}dinger equation (TDSE) with
single-active-electron (SAE), Tong and Lin also proposed a more
accurate formula \cite{Tong}, which draws up the well-known
Ammosov-Delone-Krainov ionization rate \cite{ADK} to the numerical
simulation via an empirical correction factor, thereby predicting
the ionization rate quite precisely.

Nevertheless, for double (or multielectron) ionization, the
underlying dynamics become more complicated and have not been
fully understood so far. It is generally recognized that two
electrons can be ionized either in a non-sequential or sequential
process \cite{review}. In the former mechanism, so-called
non-sequential double ionization (NSDI), the $e$-$e$ correlation
plays an essential role and the NSDI yield is remarkably enhanced
compared with the independent-electron ionization theory
\cite{Walker}. In the latter mechanism, so-called sequential double
ionization (SDI), two electrons are released independently step by
step, therefore the SDI yield can be simulated by using ionization
theory based on SAE. Nevertheless, recent experiments
\cite{SDI_exp,Keller2,Fleischer} have doubted the validity of
independent-electron approximation in the SDI regime.
Specifically, Fleischer {\it et al.} \cite{Fleischer} reported an
angular correlation between the first and second ionization
steps. In Ref. \cite{SDI_exp}, Pfeiffer {\it et al.} measured
the release time of SDI for Ar by using the attoclock technique
\cite{Attoclock}. The measured release time is in good agreement
with the prediction of Tong-Lin's formula for the first electron,
but is much earlier than the theoretical prediction for the second
electron. Such a finding has attracted significant interests immediately. Zhou
{\it et al.} \cite{Zhou} have reproduced the experimental release
time with a classical model by including the $e$-$e$ correlation.
Later, Wang {\it et al.} \cite{XuWang} argued that the release
time can be successfully explained with a similar classical model
by properly adopting the soft-core parameter but excluding the
$e$-$e$ correlation, which implies that $e$-$e$ correlation is not
essential in SDI.

Even though the classical model is robust, which has been
demonstrated extensively in previous investigations
\cite{Wang2,Uzer,Ho,XLiu}, and also provides a straightforward
interpretation with electron trajectory analysis
\cite{Zhou,XuWang}, the physical picture of SDI is still unclear.
Actually, to produce the experimental data, a scaling factor is
adopted in \cite{Zhou,XuWang} to shift the simulation curve.
Strictly speaking, the experiment has not been reproduced. Several
curial questions remain: Does the $e$-$e$ correlation play an
essential role in SDI? What is its influence and how can its
effect be observed? Can we find another indicator to identify the
$e$-$e$ correlation, except for ionization times and angular
correlation \cite{Fleischer,SDI_exp}? To clarify these questions,
we investigate SDI of Ar in an elliptically polarized field by
using a quantum model. We show that the $e$-$e$ correlation does
affect SDI. It leads to that a small part of the electrons
ionize in a correlated way. This effect
influences the observed ionization time only slightly, but it affects the overall shape of the momentum
distributions.

A complete description of a two-electron system in an elliptically
polarized field requires to simulate the motion for each electron
in at least two dimensions, which is almost impossible at present
\cite{WBecker}. To overcome the enormous computational challenge,
the so-called crapola quantum model \cite{Crapora} was introduced,
which has excellently reproduced the ``knee'' structure of NSDI
\cite{Walker,Gillen}. In this model, the total wave function is
written,
\begin{equation}
\Psi(\mathbf {r}_1,\mathbf {r}_2,t)={\psi_1}({\mathbf
r}_1,t)\psi_2(\mathbf
{r}_2,t)+\psi_2(\mathbf{r}_1,t)\psi_1(\mathbf {r}_2,t),
\end{equation}
In the high-intensity elliptically polarized field, the electron is quickly removed from the core and the recollision is significantly suppressed. Then we can assume that the overlap between the two electrons is
sufficiently small for the exchange term and can be negligible in
comparison to the Coulomb repulsion. Thus only the
first term needs to be considered. The time evolutions of the first (i.e., outer
electron, $\psi_1(\mathbf {r}_1,t)$) and second (i.e., inner
electron, $\psi_2({\mathbf r}_2,t)$) electrons are described by
\begin{equation}
i\frac{\partial \psi_n(\mathbf {r}_n,t)}{\partial
t}=[-\frac{\nabla^2}{2}+V_n(\mathbf {r}_n,t)+V_{\rm int}(\mathbf
{r}_n,t)]\psi_n(\mathbf {r}_n,t),
\end{equation}
where atomic units (a.u.) are adopted and $\mathbf {r}_n$ denotes
two-dimensional coordinates $(x_n, y_n)$, $n=1, 2$ refer to the
first and second electrons, V$_{\rm int}=\mathbf
{r}_n\cdot~{\mathbf E}(t)$ and ${\mathbf
E}(t)=E_0\exp(-t^2/\tau^2/2)[\epsilon/\sqrt{\epsilon^2+1}\cos{(\omega~t+\phi_0)}\hat{\mathbf{x}}+1/\sqrt{\epsilon^2+1}\sin(\omega~t+\phi_0)\hat{\mathbf{y}}]$
is the driving laser field. $\epsilon$, $\tau$ and $\phi_0$ refer
to the ellipticity, duration and carrier-envelope phase (CEP),
respectively. In the crapola quantum model, the first electron was
assumed to move in a static effective potential due to the nucleus
and inner electrons, i.e.,
$V_1(\mathbf{r}_1,t)=-1/\sqrt{\mathbf{r}_1^2+a_1}$. For the second
electron, the interaction with the first electron was taken into
account. Thus, V$_2$ includes a static effective potential due to
the nucleus plus a time-dependent potential due to the first
electron,
\begin{equation}\label{e3}
V_2(\mathbf{r}_2,t)=\frac{-2}{\sqrt{a_2+\mathbf{r}_2^2}}+\int
d\mathbf{r}_1
\frac{\alpha\psi_1^*(\mathbf{r}_1,t)\psi_1(\mathbf{r}_1,t)}{\sqrt{\beta+(\mathbf{r}_1-\mathbf{r}_2)^2}}.
\end{equation}
where $a_1$, $a_2$ and $\beta$ are the soft-core parameters for
the e-ion and $e$-$e$ interactions. The second electron is correlated
with the first one through the time-dependent term of Eq.
\ref{e3}. 
We could include or exclude the $e$-$e$ correlation by setting
$\alpha=1$ or 0. In the former case, we adjust the $e$-$e$
correlation strength by changing $\beta$ to identify its
influence. Note that in the crapola quantum model, the
correlation of the first electron on the second is taken
into account while its counteractive, i.e., the time-dependent
potential of the second electron on the first electron, is not
explicitly included. In order to remedy this seemingly
unreasonable treatment, we performed other calculations where the
time-dependent potential of the second electron on the first
electron is also explicitly taken into account. Thus, the
potential for the first electron is
\begin{equation}\label{e4}
V_1(\mathbf{r}_1,t)=\frac{-1}{\sqrt{a_1+\mathbf{r}_1^2}}+\int
d\mathbf{r}_2
\frac{\alpha\psi_2^*(\mathbf{r}_2,t)\psi_2(\mathbf{r}_2,t)}{\sqrt{\beta+(\mathbf{r}_1-\mathbf{r}_2)^2}}.
\end{equation} We call this method revised correlation model in this work.

We employ the split-operator method \cite{MFeit} to numerically
solve Eq. (2). Like in Ref. \cite{XTong}, the two-dimensional
space of each electron is partitioned into two regions: the outer
region A \{$|{r_n}|>a$\} and the inner region B \{$|{r_n}|<a$\}
with $a$=100 a.u. In the inner region, the wave function is
propagated exactly in the presence of combination of the Coulombic
potential and the laser field. In the outer region, the wave
function is propagated under the Volkov Hamiltonian analytically
and the final momentum spectra of the first electron $C_1(\mathbf
{p}_1)$ and the second $C_2(\mathbf {p}_2)$ are obtained from the
wave function in this region \cite{XTong}. The inner and the outer
regions are smoothly divided by a splitting technique\cite{XTong}.
The ground states of the two electrons are obtained by
imaginary-time method. In order to identify how $e$-$e$
correlation affects SDI, calculations were performed by treating
the time-dependent potential of $e$-$e$ interaction in different
ways as mentioned above, i.e., including (revised correlation
model) and excluding (uncorrelated model) it for both electrons,
and the crapola model. The data shown below are the results from the uncorrelated
and revised correlation model, while the results from the crapola
model is shown in the supplemental material for comparison
\cite{Supplemental}.

\begin{figure}[h]
\includegraphics[width=10cm,clip]{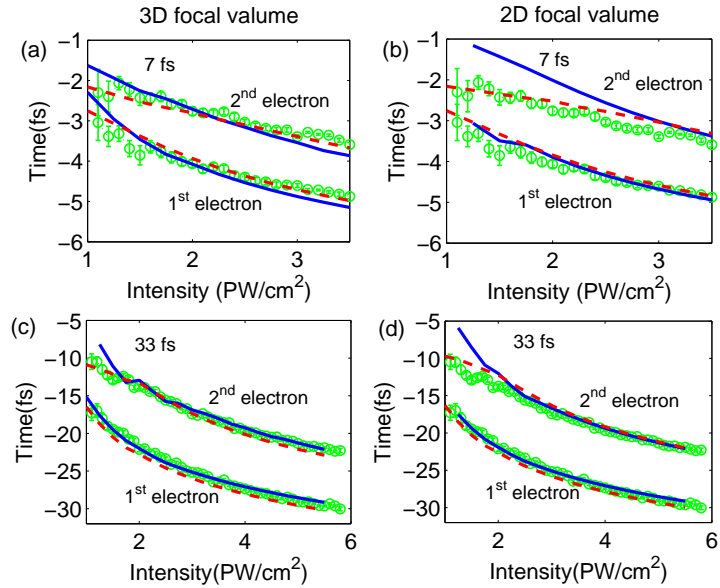}
\caption{(Color online) Release times of the first and second
electrons. (a) and (b) for the 7-fs pulse, (c) and (d) for the
33-fs pulse. The peak of pulse is at time zero. The left and right
columns correspond to 3D and 2D focal volumes. Circles with error
bars are experimental data \cite{SDI_exp}. Solid lines show the
simulation results using uncorrelated model,i.e., $\alpha=0$, $a_1=0.4$,
$a_2=1.0$. Dashed lines show the simulation result
with revised correlation model, i.e., $\alpha=1$, $\beta=4$, $a_1=0.18$,
$a_2=1.04$. } \label{fig1}
\end{figure}

Figure \ref{fig1} shows the simulated ionization times of the
first and second electrons for the 7-fs and 33-fs pulses. The
central wavelengths of the 7-fs and 33-fs pulses are 740 nm and 780
nm and the ellipticities are 0.78 and 0.77, respectively. For the
7-fs pulse, four CEP values, 0, $\pi$ and $\pm \pi/2$, are adopted
and the final simulation results are averaged over the CEP. For a
comparison with experimental data, the intensity profile of the
laser focus, the density distribution of the atoms in the gas jet,
and the geometrical overlap of the focus and the gas jet need to
be considered. Because of uncertainties of these parameters in
experiment \cite{SDI_exp}, a 3D focal volume averaging (assumption
of constant gas density across the entire 3-dimensional Gaussian
beam \cite{focal}) and a 2D focal volume averaging (assumption of
constant gas density only at the waist of the beam and zero gas
density otherwise, indicating a gas jet small compared to the
Rayleigh range of the laser focus) are tested here. We first
analyze the result for the 3D focal volume averaging. As shown in
Fig. \ref{fig1}(a), the simulated ionization time agrees well with
the experimental data for both electrons, regardless of the
$e$-$e$ correlation being excluded (solid lines) or included
(dashed lines). For the 2D focal volume averaging, as shown in
Fig. 1(b), both calculations also agree very well with the
experimental data for the first electron. For the second electron, the
calculations with uncorrelated model agree with the
experimental data at the high laser intensity range while deviates
slightly at the low laser intensity range. The calculations with revised correlation model agree better with the
experimental data at low intensity.

Figure \ref{fig1}(c) and (d) show the simulation results for
the 33-fs pulse by assuming a 3D and 2D focal configuration,
respectively. One can see that the $e$-$e$ correlation indeed
influences the ionization time, but the influence is slight. Additionally, the absolute values of the
ionization times depend also on the details of the focal geometry,
which is unknown in experiment. The ionization times simulated by
uncorrelated model, revised correlation model and crapola model (see Fig. S2 in \cite{Supplemental}) all can explain the experiment if accounting for the uncertainties in experiment. 
Thus, it is difficult to judge the
role of $e$-$e$ correlation in SDI based the ionization time.
These ionization times read in the experiments and our
calculations are integral signals, which have erased the details
in SDI, obstructing our way on revealing the role of $e$-$e$
correlation in SDI. Therefore, another indicator which keeps more
details of the ionization process and robust against
uncertainties in the focal volume averaging, is needed to identify
the role of $e$-$e$ correlation.



\begin{figure}[h]
\includegraphics[width=8.5cm,clip]{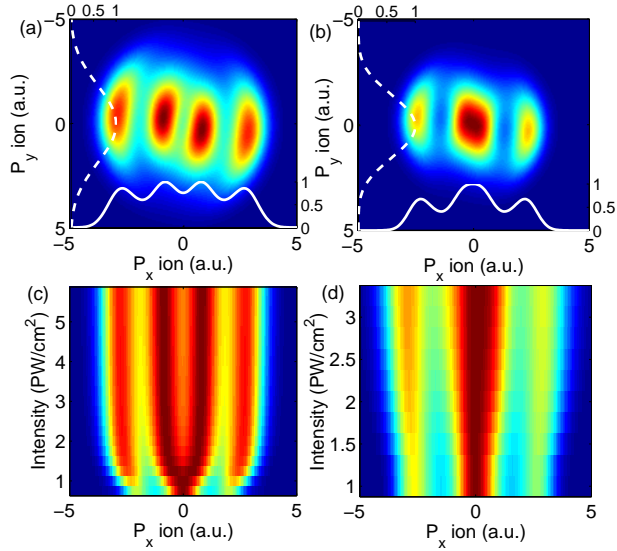}
\caption{(Color online) Two-dimensional momentum distribution for
33-fs pulse at high [(a): 3 PW/cm$^2$] and low [(b): 1 PW/cm$^2$]
intensities. The solid and dashed lines show the projection onto
the x and y axis. (c) and (d): Momentum distribution of Ar$^{2+}$
ion along the x axis as a function of laser intensity for the
33-fs and 7-fs pulses. Electron correlation is taken into account
and 3D focal volume averaging is applied.} \label{fig2}
\end{figure}

Figure \ref{fig2}(a) and (b) show the two-dimensional momentum
distribution of Ar$^{2+}$ for 33-fs pulse at 3 PW/cm$^2$ and 1
PW/cm$^2$, respectively. Here the $e$-$e$ correlation is taken
into account by using the revised correlation model and 3D focal volume configure is considered.
By projecting onto the major axis of the ellipse (y axis), the
momentum distribution is close to Gaussian, whereas along the
minor axis (x axis), it displays 3 peaks at low intensity and 4
peaks at high intensity. The outer two peaks correspond to
electrons that are emitted into parallel direction and the inner
peaks correspond to antiparallel electron emission
\cite{SDI_exp,Maharjan}. Figure \ref{fig2}(c) shows the momentum
distribution along the minor axis as a function of laser
intensity. As shown in this figure, the inner band gradually
expands with increasing the intensity and bifurcates into two
branches from 1.5 PW/cm$^2$. While for the 7-fs pulse, as shown in
Fig. \ref{fig2}(d), the inner band also expands with increasing
the intensity but does not bifurcate over the intensity range
considered in this work. All these features agree well with the
experiment \cite{SDI_exp}. Simulations are also performed by
using the uncorrelated model and crapola model. The ion momentum
distributions do not display remarkable difference except that the
depth of valley between the inner and outer peaks change slightly (see
Figs. S3 and S4 in \cite{Supplemental}).

\begin{figure}[h]
\includegraphics[width=8.5cm,clip]{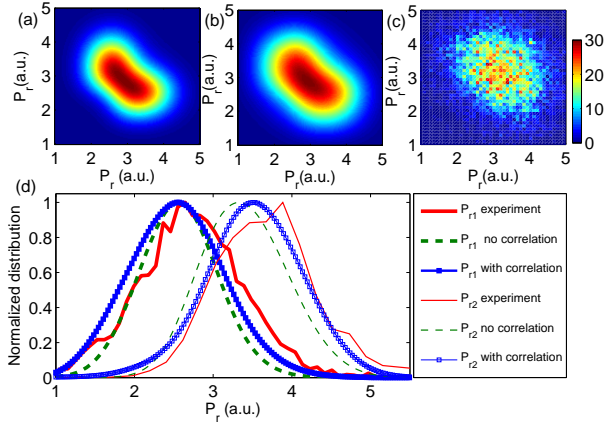}
\caption{(Color online) Correlation spectrum of radial momentum
P$_{r1}$ and P$_{r2}$ for the 7-fs pulse with peak intensity 3
PW/cm$^2$. Uncorrelated model and revised correlation model are used in (a) and (b). (c) experimental
correlation spectrum. (d) reduced spectrum of P$_{r1}$ and
P$_{r2}$. Here, the 3D focal volume averaging is applied.}
\label{fig3}
\end{figure}

To get a deeper insight into the role of $e$-$e$ correlation, we
investigate the correlation spectrum of the radial momentum, which
is defined as
P$_{r}=\sqrt{[(\epsilon^2+1)/\epsilon^2]P_x^2+(\epsilon^2+1)P_y^2}$,
and P$_{r1}$ and P$_{r2}$ are the radial momenta of the first and
the second electrons, respectively. Figures \ref{fig3}(a) and (b)
show the simulation results by using the uncorrelated model and
the revised correlation  model for the 7-fs pulse. Figure
\ref{fig3}(c) presents the correlation spectrum observed in the
experiment. The difference between Figs. \ref{fig3}(a) and (b) is
remarkable. The distribution including $e$-$e$ correlation is
fatter than that excluding $e$-$e$ correlation. The result
including $e$-$e$ correlation agrees better with the experimental
observations. This feature can be more clearly seen in the reduced
spectra, which are respectively obtained from the momentum spectra
of the first electron $C_1(\mathbf {p}_1)$ and the second electron
$C_2(\mathbf {p}_2)$. As shown in Fig. \ref{fig3}(d), for both
electrons, the widths of the spectra excluding $e$-$e$ correlation
are narrower than the experimental data. While for the spectra
including $e$-$e$ correlation, the widths are in agreement with the
experimental data. In order to further confirm that the broadening
of the reduced spectra originates from the $e$-$e$ correlation, we
performed other calculations where the strength of $e$-$e$
correlation is enhanced by decreasing the soft-parameter $\beta$.
As shown in the supplementary material \cite{Supplemental}, the reduced spectra become wider for both
electrons for a smaller value of $\beta$. Note that a fat radial momentum distribution is also obtained by the crapola model \cite{Supplemental}. These results indicate that SDI indeed is
influenced by $e$-$e$ correlation and this influence leaves
imprint on the electron momentum spectra.

As shown in Figs. 3(a) and 3(b), the peaks of the momentum
spectra do not change dramatically when $e$-$e$ correlation is
excluded or included. Therefore the mean value of radial momentum
is not dramatically influenced by $e$-$e$ correlation. In the
attoclock experiment \cite{SDI_exp}, the ionization time is
extracted from the radial momenta. Therefore, the influence of the
$e$-$e$ correlation on the averaged ionization time is not very
effective. Similar remarks also can be observed for the 33-fs
pulse. This feature also explains the puzzle why the ionization
time can be reproduced with the classical model either including
or excluding the $e$-$e$ correlation \cite{Zhou,XuWang}.


\begin{figure}[h]
\includegraphics[width=6cm,clip]{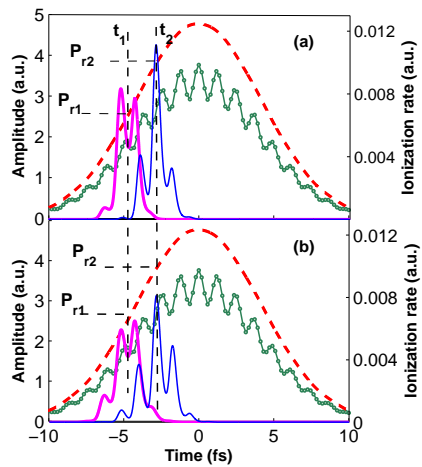}
\caption{(Color online) Ionization rate of the first (bold magenta
lines) and second (thin blue lines) electrons as a function of
time for the 7-fs pulse with peak intensity of 3 PW/cm$^2$. Uncorrelated model and revised correlation model are adopted in (a) and (b).
The solid lines with dots and the dashed lines
represent the vector potential of the laser field and the
radial momentum, respectively} \label{fig4}
\end{figure}

The origin of the broadening of the reduced momentum spectra is
related to the time intervals when the first and the second
ionizations occur. Figure \ref{fig4}(a) shows the ionization rate
as a function of time at 3 PW/cm$^2$ for the 7-fs pulse. The bold and
thin solid lines correspond to the first and second electrons
without $e$-$e$ correlation. Because the laser intensity is much
higher than the saturation intensity, the ionization is confined
in a single cycle around t$_1$ and t$_2$ at the ascending part of
the pulse, respectively. After
ionization, the electrons are accelerated by the laser field and
finally released with a momentum of P$_{r1}$ and P$_{r2}$ as shown
in Fig. \ref{fig4}(a). But because of no correlation, the second
electron is not freed when $t<t_1$ and the first ionization process is already
finished when $t\ge t_2$. In contrast, when the $e$-$e$
correlation is accounted for [Fig. \ref{fig4}(b)], the centra of
the ionization time ($t_1$ and $t_2$) for both electrons
almost do not change. This is consistent with the above results
that the ionization time is not effectively influenced by the
$e$-$e$ correlation. However, as shown in Fig. 4(b), when $t<t_1$,
small percentage of the second electron starts to be released and also
small percentage of the first electron is released at $t\ge t_2$.
Consequently, the ionization windows of the first and second
ionization processes become broader than those excluding the
$e$-$e$ correlation. The long ionization window broadens the
width of the momentum distribution. Moreover, because the first
electron quickly moves away after ionization, its correlation with
the second electron drops quickly. Only a small percentage of the first
and second electrons can be ionized in a correlated way. By
increasing the $e$-$e$ correlation, larger percentage of the
electrons can be ionized in a correlated way and the momentum
spectra will be further broadened as shown in the supplementary material
\cite{Supplemental}. Therefore, the width of the momentum
spectrum, which is closely related to the ionization window,
exhibits the faint details of $e$-$e$ correlation in SDI.

In conclusion, SDI of Ar subjected to elliptically polarized
pulses is investigated using quantum models. It is shown that the $e$-$e$ correlation indeed slightly shifts the ionization
time, but its influence is not crucial to explain the experiment \cite{SDI_exp}. Moreover, uncertainties in the focal volume
averaging also influence the apparent ionization times. It is difficult to decisively reveal the
role of $e$-$e$ correlations based on the ionization timing data alone. However, our simulations show that the $e$-$e$
correlation broadens the time windows for the first and second
ionizations. Though this effect could not be distinguished in the
ionization time of the second electron, it affects the momentum
distribution. The signature of $e$-$e$ correlation can be
identified in the width of the electron momentum distribution. By
comparing with the experiment \cite{SDI_exp}, the width of the
electron momentum distribution for 7-fs pulse agrees well with
the simulation including fair $e$-$e$ correlation, which indeed
indicates the exist of $e$-$e$ correlation in SDI. The subtle
features of momentum distribution require further experimental
measurements with low noise and high accuracy.

\section*{Acknowledgement}
We acknowledge helpful discussions with Dr. E. L\"{o}stedt, Dr. X. Wang and Prof. J. H. Eberly. We
thank Dr. C. Cirelli and Prof. U. Keller for kindly providing experimental data. This
work was supported by the National Science Funds
(No. 60925021, No. 61275126 and No. 11234004) and the 973 Program
of China (No. 2011CB808103).

\end{document}